Low-energy band structure in Bernal stacked hexlayer graphene: Landau fan diagram and resistance ridge

Tomoaki Nakasuga[1], Shingo Tajima[1], Taiki Hirahara[1], Ryoya Ebisuoka[1], Takushi Oka[1], Kenji Watanabe[2], Takashi Taniguchi[2] and Ryuta Yagi[1*]

[1]Graduate School of Advanced Sciences of Matter, Hiroshima University, Kagamiyama 1-3-1, Higashi-Hiroshima, Hiroshima 739-8530, Japan
[2]National Institute for Materials Science (NIMS), Namiki 1-1, Tsukuba, Ibaraki, 305-0044 Japan.

**Abstract**

 The intrinsic peaks due to the topological transition of the Fermi surface shape, which appear in the resistivity as a function of carrier density and perpendicular electric field, were studied in AB-stacked hexlayer graphene as a system with three bilayer-like bands. By the Landau level structure measured in low temperature magnetotransport experiments, and by band structure calculations, it is shown that the ridge structures correspond one-to-one with the characteristic positions in the dispersion relations. Some ridges originated from electronic states near the bottoms of the bilayer-like bands that appears at the carrier density of the zero-mode Landau levels. Some ridges originated from the mini-Dirac cones formed in the perpendicular electric field because of hybridization of bilayer-like bands.

 * Corresponding Author     yagi@hiroshima-u.ac.jp

1. Introduction

Since graphene was successfully made by cleaving bulk crystal of graphite with adhesive tape [1], a number of experimental and theoretical investigations have been performed to elucidate physical property and its potential applications. Band structure of graphene is expected to vary depending on the number of layers and stacking [2-9]. In case of AB-stacked graphene it can be understood via "parity effect" with respect to the number of layers [2-7]: in the case of an even number of layers, *i.e.*, $2n$ ($n = 1,2,3 ...$), the electronic band structure consists of $n$ bilayer-like bands, and in the case of an odd number of layers, *i.e*, $2n + 1$ ($n = 1, 2, 3 ...$), it consists of a monolayer-like band and $n$ bilayer-like bands. To date, the band structure has been experimentally studied by using optical [10-17], and transport methods [18-28].

The electronic band structure of graphene can be tuned by the electrostatic potential arising from perpendicular electric fields [18,22,26-27,29-30] created by top and bottom gate voltages. They open an energy gap in the bilayer-like bands [15,31-36] and bilayer graphene becomes insulating. On the other hand, AB-stacked trilayer graphene does not show insulating behavior [35,37-40]; although an energy gap appears at the bottom of the bilayer band, the density of states arising from an extra monolayer-like band. Recently, investigations of the response of the band structure to a perpendicular electric field uncovered a new feature of AB-stacked tetralayer graphene. It showed resistance ridges in the top and bottom gate voltage dependence of resistance at zero magnetic field [26-27]. These resistance ridges are interpreted to be an intrinsic property of AB-stacked tetralayer graphene, which originates from the topological transition of the Fermi surface shape [27]. AB-stacked tetralayer graphene has two sets of bilayer-like bands which are hybridized. The conspicuous resistance ridges appear at the bottoms of the bilayer bands [26-27].

The ridges correspond one-to-one with the characteristic points in the dispersion relation, and this would enable ones to detect band structure of two-dimensional materials, as a complementary method to the quantum oscillations. Therefore, further study of the resistance ridge structure is required to reveal the nature of the resistance ridges. In this paper, we studied resistance ridge structures in AB-stacked hexlayer graphene. Since AB-stacked hexlayer graphene is graphene with the smallest number of layers that has three sets of bilayer-like bands, as shown schematically in Fig. 1 (a), we can investigate a new case of the resistance ridges due to topological transition, which is much complicated than the case of AB-stacked tetralayer. We will show that ridge structure results not only from the bottoms of bilayer-like bands, but also from gap structures, some of which form mini-Dirac cones.

## 2. Experimental

Our sample consisted of graphene encapsulated with thin *h*-BN flakes, on which were formed top and bottom gate electrodes. Flakes of graphene and h-BN were prepared by mechanical exfoliation of high-quality bulk crystals. An optical micrograph of the sample and a schematic diagram of the sample structure are shown in Figure 1 (b) and (c). A flake of moderately thick graphene, which was transferred onto the top of the stacks, served as a top gate electrode. The $SiO_2$ substrate, which was heavily doped and conducting at low temperature, served as a back gate electrode. Electrical contacts with encapsulated AB-stacked hexlayer graphene were formed using the edge contact technique [41]. Electrical contacts with graphene serving as the top gate were formed at positions on the substrate where the encapsulated AB-stacked hexlayer graphene was absent.

Identifications of the number of layers and stacking are one of the most important points in this study. The number of layers was identified by combining information of on the optical contrast in the graphene image, atomic force microscopy (AFM), and the shape of the G' band of the Raman spectra. The stacking was verified by examining the shape of the Raman G' band and the structure of the Landau fan diagram (see Appendix 2 for details).

The magnetoresistance measurements were performed at $T = 4.2$ K. Magnetic fields were applied using a superconducting solenoid. The resistivity measurements were done using the standard lock-in technique.

## 3. Results and discussions

### 3.1 Resistivity at zero magnetic field

Back gate voltage ($V_b$) dependence of resistivity exhibited a significant dependence on the top gate voltage ($V_t$) as shown in Fig. 2 (a). The resistivity traces show conspicuous peak structures which are due to the charge neutrality. The peak resistivity increases with increasing $|V_t|$, which would be relevant to insulating behavior arising from the perpendicular electric field, as in bilayer graphene [32,38,40]. This is apparently different from that of AB-stacked trilayer, which shows the opposite behavior [38,40]. In addition to the large peak, small peaks are discernible in each trace in Fig. 2 (a). The overall structure is reminiscent of recent experiments on the AB-stacked tetralayer graphene that show multiple peak structures [26-27]. To study the observed peak structure further, we measured the detailed response of the resistivity $\rho$ to $V_b$ and $V_t$. Figure 2 (b) shows a map of resistivity as a function of $V_b$ and $V_t$, which were measured at $B = 0$ T and $T = 4.2$ K. $\rho$ shows a significant variation with the top and bottom gate voltages. The peak resistivity, which is approximately on a straight line that passes through the origin is the trace of the main peak in Fig. 2(a), which is due to charge neutrality. In addition, a curved ridge structure, which looks like a set of hyperbolas, is discernible. Similar structures were observed in AB-stacked tetralayer graphene [26-27].

Applying gate voltages to the top and bottom gate electrodes allowed us to control the perpendicular electric fields and carrier density independently. The top gate voltage $V_t$ induces charge density, $-C_{tg}(V_t - V_{t0})$, in graphene, while the back gate voltage $V_b$ induces charge density, $-C_{bg}(V_b - V_{b0})$. Here, $C_{tg}$ and $C_{bg}$ are the specific capacitances of the top and bottom gate electrodes, respectively. $V_{t0}$ and $V_{b0}$ are variation of the neutrality point due to the offset charge arising from, for example, substrates, and were estimated to be 1.02 and 1.49 V, respectively. The total carrier density $n_{tot}$ is $(C_{tg}(V_t - V_{t0}) + C_{bg}(V_b - V_{b0}))/(e)$. The electric flux density perpendicular to the sample is a half of the difference of the charge density induced by bottom and top gate voltage, *i.e.*, $D_\perp = (C_{tg}\ V_t\ (V_t - V_{t0}) - C_{bg}(V_b - V_{b0}))\ /\ 2$ [42]. The electric field in graphene perpendicular to the plane can be written as $E = D_\perp/(\epsilon\epsilon_0)$ where $\epsilon_0$ is the dielectric constant of vacuum if the relative dielectric constant of graphene is given by a constant value $\epsilon$.

The condition for constant $D_\perp$ can be attained by varying the top and bottom gate voltages to satisfy the condition, $(V_b - V_{b0}) = C_{tg}/C_{bg}\ (V_t - V_{t0}) + V_0$, where $D_\perp = -C_{gb}V_0/2e$. On the other hand, the ratio of the capacitance $C_{tg}/\ C_{bg}$ could be calculated from the conspicuous ridge structure for the charge neutrality in Fig. 2 (b). Here, the top gate and bottom gate voltage satisfy the condition, $C_{tg}\Delta\ V_t = -C_{bg}\Delta\ V_b$. This gives the ratio, $C_{tg}/\ C_{bg} \approx 7$ ( $C_{bg} = 107$ aF/μm$^2$ ).

Figure 2 (c) shows a replot of Fig.2 (b) as a function of total carrier density $n_{tot}$ and $D_\perp$. The ridge structure for charge neutrality is vertically elongated at $n_{tot} = 0$. In addition, hyperbolic ridges appear symmetrically in $D_\perp$.

### 3.2. Landau level structures

Because the resistance ridge structure would be relevant to the electronic band structure, we took the same methodology as Ref. [28] to identify the origin of each ridge structures. We studied the Landau level structure in detail, from which we deduced the low-energy band structures at a zero magnetic field. The magnetoresistance was measured as a function of the top and bottom gate voltages. Figure 3 (a) shows a map of longitudinal resistivity $\rho_{xx}$ as a function of $n_{tot}$ and magnetic field *B* for zero perpendicular electric fields, *i.e.*, $D_\perp = 0$, which minimizes variation of internal electrostatic potential. Shubunikov-de Haas (S-dH) oscillations manifest themselves as bright and dark patterns. Unlike mono- and bilayer graphene [1], the Landau levels are not simple fan-shaped structures, but possibly consist of overlaid fan-shaped structures that have complicated level crossings [18-22,26-28,43]. This complexity arises from the fact that the low-energy band structure of the AB-stacked hexlayer graphene consists of three bilayer bands

[3,5-8,22,44].

One of the most characteristic features in Fig. 3 (a) is the emergence of zero-mode Landau levels $\alpha$, $\beta$, and $\gamma$, which are located near the charge neutrality point. Each of these zero modes remains approximately at the same value of $n_{tot}$ as $B$ increases. The degeneracy of the zero-mode Landau levels is a clue to determining the nature of each band. The degeneracy can be identified by inspecting the variation in the filling factors of the energy gaps associated with the Landau level crossings. Some of the energy gaps and filling factors are indicated on the Landau fan diagram in Fig. 3 (b). It is easy to verify that the zero-mode Landau levels $\alpha$ and $\gamma$ have eight-fold degeneracy, which is a characteristic feature of bilayer-like bands [1,45-47]. The zero-mode Landau level $\alpha$ is between gaps with $\nu = -8$ and $-16$ at $B \approx 4$ T, and between $-12$ and $-20$ at $B \approx 3$ T. The zero-mode $\gamma$, on the other hand, is between gaps with 8 and 16 at about $B = 6$ T and between gaps with 12 and 20 at $B \approx 5$ T. (The relevant positions are indicated by horizontal lines.) The difference between the filling factors of the gap is eight, which is the degeneracy of the zero-mode Landau levels of the bilayer-like bands. This contrasts with the case of a monolayer-like band. The zero-mode Landau level has a degeneracy of four [1,19,21-22].

To better understand the experimentally obtained fan diagram, we performed theoretical Landau level calculations, from which we could determine band parameters required for calculating dispersion relation at zero magnetic fields. In addition, it could be confirmed that the measured graphene is AB-stacked hexlayer. The Hamiltonian is based on the effective mass approximation of the tight binding model. Figure 3 (c) shows the energy eigenvalues of the Hamiltonian of AB-stacked hexlayer graphene in a magnetic field calculated for the Slonczewski-Weiss-McClure parameters of graphite, and Fig. 3(d) is a numerically calculated Landau fan diagram, which shows density of state (DOS) as a function of carrier density. Better agreement might be obtained by fine tuning of the SWMcC parameters; however, Fig. 3(d) nonetheless reproduces important features of the Landau fan diagram of AB-stacked hexlayer graphene. In particular, the presence of three zero-mode Landau levels with eight-fold degeneracy is readily seen as vertically elongated Landau levels. In Fig. 3 (a), although the zero-mode Landau levels $\alpha$ and $\gamma$ are clearly related to the bilayer-like band, the degeneracy of the zero-mode Landau level $\beta$ is not clear as a single level because of the large resistance that grows in the vicinity of $n_{tot}$=0 where $\beta$ is expected to appear. As can be seen in Fig. 3 (c), near $\beta$, there is a four-fold degenerate Landau level that is hole-like in low magnetic fields but is electron-like at about $B > 5$ T. These two Landau levels were not resolved in the experiment and appeared between gaps with $\nu = -4$ and $+8$ in the high magnetic field regime. The fact that zero-mode Landau levels appeared at three different carrier densities shows the semi-metallic features of the AB-stacked hexlayer graphene, because zero-mode Landau levels appear at the bottoms of the conduction and valence bands. The dispersion relation in the direction of stacking, or the offset energy of zero-mode Landau levels is principally due to $\gamma_2$ of the SWMcC parameters, which represents the interlayer coupling [9,43].

So far we have studied the Landau fan diagram for $D_\perp = 0$. Next, we examine the case of $D_\perp \neq 0$. Figure 4 (a) and (b) shows maps of longitudinal magnetoresistivity measured for $D_\perp = \pm\ 0.65 \times\ 10^{-7}$ cm$^{-2}$As. Here, top and bottom gate voltages were varied so that only $n_{tot}$ was varied while keeping $D_\perp$ constant. It is clear that the zero-mode Landau levels split via the perpendicular electric field; the zero-mode Landau level $\gamma$ in the electron regime splits into two levels, $\gamma_1$ and $\gamma_2$, although the splitting is not clear in the hole regime. Splitting of zero-mode Landau levels of the bilayer-like band due to perpendicular electric fields also occurs in graphene with different numbers of layers [18,27-28].

### 3.3. Resistance ridge structures
Now we return to the resistance ridge structures, and discuss it in the light of the electronic band structure of the AB-stacked hexlayer graphene. More fine structures are found in the derivative of resistivity with respect to $n_{tot}$, as shown in Fig. 5 (a). Conspicuous ridge structures are labeled by t-z. The overall structure of the ridges is much more complicated than in the AB-stacked tetralayer graphene [26-27]. Some of the ridges can be readily identified. Ridge t originates from the neutrality point. Ridges u, v and z are due to the bottoms of bilayer like-bands, which was confirmed by the positions of zero-mode Landau levels as shown in Fig 5(c). On the other hand, ridge x appears near the zero-mode Landau level β in the vicinity of the charge neutrality point.

More detailed discussions on the origin of the ridges could be done with a help of numerical calculations of the dispersion relations. Figure 5 (b) shows the dispersion relation as a function of wave number along the $x$-direction ($k_x$), under the condition, $k_y = 0$. The Slonczewski-Weiss-McClure (SWMcC) parameters of graphite were used in the calculation, the same as those to make the Figs. 3 (c) and (d). The carrier densities were calculated from the area of the Fermi surface, *i.e.*, the energy-contour of the dispersion relations. The dispersion relation at the zero perpendicular electric field is rather complex, because complicated energy gaps are formed by mixing of bilayer-like bands and by trigonal warping. We examined carrier densities for representative points in the dispersion relations labeled by a-f and $\Delta_1$-$\Delta_2$ in Fig. 5 (b). In Fig. 5 (c), the numerically calculated $n_{tot}$ for the representative positions are plotted for different values of $|D_\perp|$. Ridges u, v, x and t correspond to the cases a, b, c and d. Broad ridge z originates from cases e and f, which were not resolved clearly as in the case of the zero-mode Landau level. These ridges are for flat-band-like structures which are formed at the bottoms of the bilayer-like bands.

Other ridges w and y are for cases $\Delta_1$ and $\Delta_2$, where mini Dirac cones are formed because of diminished or vanishing energy gaps. Three identical mini-Dirac cones are formed for each K and K' valley. The mini-Dirac cones are expected to appear in AB-stacked trilayer graphene under a perpendicular electric field [48]. The present mini-Dirac cones in AB-stacked hexlayer graphene originate from the same mechanism as in AB-stacked trilayer in that trigonal warping locally closes the energy gap formed by the perpendicular electric field. The mini-Dirac cone is known to appear also in AB-stacked tetralayer graphene, as a peak at the charge neutrality point [26,27]. On the other hand, those in hexlayer graphene are expected to appear not only near the charge neutrality point (near point d in Fig. 5 (b)) but also at other carrier densities; they appear as sharp and unsplit resistance ridges.

Band structure of AB-stacked hexlayer graphene is highly complicated because of trigonal warping. However schematic structure, *e. g.*, energies for the bottoms of the bilayer-like bands, are principally determined by $\gamma_2$ in the SWMcC parameters of the tight-binding model. The detailed structures including the mini-Dirac cone, are modified by $\gamma_3$ which results in trigonal warping. Simple dispersion relation calculated without considering trigonal warping would allow us to understand the schematic structure of the resistance ridge. Figure 6 (a) and (b) show similar plots as Fig. 5 (b) and (c), which displays result calculated for the same SWMcC parameters, excepting that $\gamma_3$ was set to zero. As seen in the figure, the representative positions in the band, which were labelled by a-f and $\Delta_1$-$\Delta_2$ in Fig. 6 (a) reproduces experimentally observed ridge structures. This indicates that carrier densities of the representative points in the dispersion relations are approximately unchanged by $\gamma_3$, at least in low energy regimes relevant to this transport measurements. Energy gaps between $\Delta_1$ and $\Delta'_1$, or between , between $\Delta_2$ and $\Delta'_2$, which are isotropic in this calculation, originate from hybridization of bilayer-like bands. The mini-Dirac cones are associated with these gap structures, which becomes anisotropic via $\gamma_3$. Because electronic states near the band gaps that vary with $\gamma_3$ are rather limited, the resistance ridges appears approximately at the same positions.

The band structure of graphene becomes more complicated as the number of layers increases, and accordingly, the band mass varies complicatedly in the reciprocal space. Phenomena due to this complication should appear in the resistivity measured at zero-magnetic field. A detailed measurement of resistivity at zero magnetic field might be another method for studying the electronic band structures of graphene, or two-dimensional materials, by transport measurements.

## 4. Conclusion

The resistance ridge structures arising from the topological transition of the Fermi surface shape were studied in AB-stacked hexlayer graphene in terms the band structure which was determined by magnetotransport measurements.using samples with top and bottom gate electrodes. Some of the ridge structures were identified to originate from the bottoms of the conduction and valence bands of the bilayer-like band, which is closely related with $\gamma_2$ in SWMcC parameters. We also found other ridges that stemmed from energy gap structure associated with hybridization of bilayer-like bands. This gap structure is expected to form mini-Dirac cones in perpendicular electric fields due to trigonal warping.

# Appendix:

## 1. Sample Fabrication

Graphene samples were prepared by mechanically exfoliating high-quality Kish graphite. A graphene flake was encapsulated with $h$-BN flakes prepared by using mechanical exfoliation of a high-quality $h$-BN crystal. The encapsulation technique is described in Ref. [41]. To make a dual-gated sample, the encapsulated graphene was formed on heavily doped Si substrate covered with $SiO_2$. The Si substrate remains conductive at $T$ = 4.2 K, and hence serves as a bottom gate electrode. The top gate electrode was formed from a few-layer graphene flake, which was transferred onto the top of the encapsulated graphene by using the technique described in Ref. [19]. Samples for the measurement were made by electron beam lithography, ion etching, and the lift off technique. First, the top gate (a few layer graphene) was defined by a reactive ion etching using a mixture of low pressure $CF_6$ and $O_2$ gas, so that the top gate covers an effective sample area but not the area of the contacts. The sample was then patterned into a Hall bar. In this sample structure, the top gate and the effective sample area have exact geometry. Electric contacts to the graphene were made by using the edge contact technique [41].

## 2. Determining the number of layers and stacking

The number of layers of graphene was determined by using various methods. After the mechanical exfoliation of the graphite, hexlayer graphene flakes were found by analyzing the color of the flakes [49-51]. The relationship between number of layers and the color of the graphene flakes was examined by measuring the thickness of the graphene flakes by using AFM topography. We also performed Raman spectroscopy using samples with a known number of layers. Raman G' peaks were measured for many graphene samples. Most of them revealed a systematic variation with the number of layers. There are reports on measuring the Raman G' peaks of AB-stacked graphene with 1-8 layers [22,52-57]. Figure 7 (a) shows the Raman G' spectra. Our data approximately reproduced the line shapes of the reported AB-stacked multilayer graphene [52-57]. The Raman G-band spectra also showed a systematic enhancement with the number of layers as reported in Refs. [22,52,58]. Moreover, systematic variations appeared most frequently in flakes that were exfoliated from Kish graphite, which is consistent with that the samples are AB-stacked which is the most stable. We also found flakes whose spectra were qualitatively different from the Raman G' band and whose stacking was is possibly ABC [53-54].

There is a possibility that graphene flakes have domains with different stackings, despite that the number of layers is constant in the flake. In the previous studies, the stacking sequence was detected in a Raman mapping experiment examining the ratio of spectra intensities at two different Raman shifts in the G' band spectra [55,58]. We found that mapping the spectra intensity at a Raman shift near the Raman G peak clearly revealed differences in the stacking sequence [22]. We prepared many graphene flakes from graphite crystals supplied by different vendors. Fakes with ABC-stacking were found, though rarely in some batch of graphite crystal. In other batches, ABC stacking was not found at all.

We could not perform Raman spectroscopy on the samples used in the experiment, because they consisted of encapsulated graphene with a top gate of thick graphene. That is, a Raman measurement could not be performed on hexlayer graphene underneath a thick top gate graphene. Also, a Raman mapping experiment would have been difficult to perform during the sample fabrication process used in our laboratory because it could contaminate the surface of the sample. Moreover, the thick top gate insulator (a $h$-BN flake) to prevent leakage current from the top gate made it impossible to perform the Raman spectroscopy on the hexlayer graphene beneath the thick gate insulator. For the above reasons, we performed similar experiments using other hexlayer graphene samples that had only a back gate electrode. If the Landau level structure of such a sample approximately reproduced the magneto transport results in the main text, it would confirm that the sample described in the main text was AB stacked six-layer graphene. Although graphene was also encapsulated with *h*-BN flakes in the back gate sample, the top *h*-BN flake was much thinner compared with the sample with the dual gate electrode, which enabled us to perform Raman spectroscopy. A map of the Raman G peak signal did not show any significant variation in stacking in the sample, as shown in Fig. 7(b). The shapes of the Raman G' spectrum was approximately the same as that of AB stacked hexlayer graphene displayed in Fig. 7 (a) (see Fig 7 (c)) [53-55]. A map of magneto resistivity for this hexlayer graphene sample measured at $T = 4.2$ K is displayed in Fig. 8 (a). Figure 8 (b) compares the positions of the characteristic energy gaps between the back gate sample and the sample described in the main text. The Landau fan diagrams have approximately the same structure, which proves that the number of layers and stacking were the same [22].

## 3. Calculation of Landau levels

The Landau levels were calculated using an effective mass approximation based on the tight-binding model [43]. We used a Hamiltonian considering all the SWMcC parameters, *i.e.*, $\gamma_0$, $\gamma_1$, $\gamma_2$, $\gamma_3$, $\gamma_4$, $\gamma_5$ and $\Delta'$ [43]. Here, $\Delta' = \Delta - \gamma_2 + \gamma_5$. The definitions of the SWMcC parameters are depicted in Fig. 9. In the present paper, $\gamma_0$ - $\gamma_5$ and $\Delta'$ for graphite were used (3.16 eV, 0.39 eV, 0.3 eV, 0.044 eV, 0.038 eV and 0.037 eV, respectively). The wave functions were expanded using Landau functions [43], and the eigenvalues were calculated numerically. Because the crystal structure of AB-stacked hexlayer graphene has a spatial reversal symmetry, the Landau levels for K and K' valleys are degenerated. The density of states was calculated by assuming that each Landau level had a Gaussian distribution in energy with widths proportional to the magnetic field.

## 4. Calculation of dispersion relation

The dispersion relations were calculated using the Hamiltonian derived for the effective mass approximation [3,5-6,44]. The wave functions were expressed as the plane waves, and the eigen values of the matrix elements were calculated numerically. The effect of the perpendicular electric field was taken into account by adding, to the diagonal elements of the Hamiltonian, the electrostatic potential arising from the charge distribution due to the top and bottom gate voltages. The principle of inducing charges in graphene is approximately the same as a capacitor consisting of two metal gate electrodes to which a voltage is applied. However, although in a framework of classical electromagnetism, induced charges appear on the surface of each electrode, one must consider the distribution of the induced charge within the graphene layers because graphene is an extremely thin material consisting of atomic layers. It is commonly accepted that the screening length is about a few layers[11,29,59-63]. In this study, we chose the permittivity of graphene to be $\epsilon/\epsilon_0 = 2$ and the screening length to be $\lambda = 0.43$ nm, which is approximately the same as the theoretical values [61], and used them to analyze our past experimental results [22].

If $D_\perp = en_g$, charges $-en_g$ and $en_g$ are induced at the top and bottom gate electrodes, respectively. The charge distributions due to the top and bottom gate voltages were calculated independently, as shown in Fig. 10. Here, we assumed that the sum of the induced charges $e(n_1 + n_2 + \cdots + n_6)$ equals the charge $en_g$ induced by the gate electrode. The charge distribution due to the top and bottom gate voltages was calculated by superposing the individual charge distributions. The electrostatic potential for each layer was calculated by solving the Poisson equation.

The carrier density was calculated from the area of the energy contours of the dispersion relations as a function of the two-dimensional wave number $\vec{k} = (k_x, k_y)$.

### 5. Characteristic energy gap in Landau level structure of AB-stacked hexlayer graphene.

There is an energy gap with characteristic filling factors near the charge neutrality point. An energy gap for a filling factor $\nu = -4$ is expected to appear between two zero-mode Landau levels, as seen in Fig. 3 (c). There is no Landau level crossing that close this energy gap above $B \sim 2.5$ T. In the experimental fan diagram, the energy gap with $\nu = -4$ appears as a resistance ditch. This kind of energy gap between zero-mode Landau levels forms characteristic energy-gap structures in the fan diagram in multilayer graphene. As another example, the gap with $\nu = 8$ is clearly visible near the charge neutrality point approximately above $B > 4$ T. The energy gap is located between the zero-mode Landau levels $\beta$ and $\gamma$. The filling factor of the gap is $\nu = 8$ (not $\nu = 4$), because an electron-like Landau level with a degeneracy of four exists between the zero-mode Landau levels $\beta$ and $\gamma$.

## Acknowledgements

This work was supported by KAKENHI No.25107003 from MEXT Japan.

# Figure Captions

**Figure 1**
**Characterization of Bernal-stacked six-layer graphene.**
(a) Simplified band structure of hexlayer graphene. (b) Schematic diagram of sample structure. A vertical cross-section is displayed. (c) Optical micrograph of a sample. Top-gate graphene covers over the region relevant to the measurement, except on the electric leads for the current and voltage probes. The bar shows 10 μm. TG means top gate. hBN means $h$-BN flake.

**Figure 2**
**Top and bottom gate voltage dependence of resistivity.**
(a) Bottom gate voltage ($V_b$) dependence of resistivity for different top gate voltages ($V_t$). $V_t$ was varied from 5 to -5 V in 1V steps. $T = 4.2$ K. $B = 0$ T. (b) A map of resistivity at zero magnetic field as a function of top and bottom gate voltages. $T = 4.2$ K. (c) Replot of panel (b) as a function of $n_{tot}$ and $D_\perp$ which is proportional to the perpendicular electric field.

**Figure 3**
**Magneto-transport measurement.**
(a) A map of longitudinal magnetoresistivity ($\rho_{xx}$) as a function of carrier density ($n_{tot}$) and magnetic field ($B$). $D_\perp$ was kept at zero. Arrows $\alpha$, $\beta$, and $\gamma$ indicate the positions of the zero-mode Landau levels. $T = 4.2$ K. (b) Filing factors $\nu$ of some of energy gaps and Landau level crossings are shown on the map. The horizontal line is a guide for eye to see the width of the zero-mode Landau levels. (c) Numerically calculated Landau levels for AB-stacked hexlayer graphene. SWMcC parameters of graphite were used ($\gamma_0 = 3.16$ eV, $\gamma_1 = 0.39$ eV, $\gamma_2 = -0.02$ eV, $\gamma_3 = 0.3$ eV, $\gamma_4 = 0.044$ eV, $\gamma_5 = 0.038$ eV, $\Delta' = \Delta - \gamma_2 + \gamma_5 = 0.037$ eV). (d) Numerically calculated density of states.

**Figure 4**
**Landau fan diagrams in the presence of perpendicular electric fields.**

A map of longitudinal magnetoresistivity for $D_\perp = 0.65 \times 10^{-7}$ cm$^{-2}$ As (a) and $D_\perp = -0.65 \times 10^{-7}$ As (b) as a function of $n_{tot}$ and $B$. $T$ = 4.2 K.

Figure 5

**Resistance ridge structure in $n_{tot}$ and $D_\perp$ dependence.**

(a) A map of the derivative of resistivity with respect to $n_{tot}$ as a function of $n_{tot}$ and $D_\perp$. Some of the ridge structures are labeled (u-z). (b) The dispersion relation of AB-stacked hexlayer graphene for different values of $|D_\perp|$, that were calculated numerically using the SWMcC parameters of graphite. Representative points in the dispersion relations are labeled a-f and $\Delta_1$-$\Delta_2$. (c) The same map, plotting the positions of zero-mode Landau levels, which were observed in the experiment, are plotted with red crosses. The other symbols indicate points a-f, and $\Delta_1$ in the dispersion relation (see panel (b)), which was numerically calculated from the area of the energy contour of the dispersion. Lines are guides for the eye.

Figure 6

**Resistance ridge and band structure calculation without trigonal warping.**

(a) Dispersion relations of AB-stacked hexlayer graphene for different values of $|D_\perp|$, which were calculated numerically using SWMcC parameters of graphite excepting $\gamma_3 = 0$. Representative points in the band structure are labeled a-f, and and $\Delta_1$, $\Delta_1{}'$, $\Delta_2$ and $\Delta_2{}'$ (b) The calculated positions for the representative points are overlaid on the experimentally determined map.

Figure 7

**Raman spectra of Bernal stacked six-layer graphene.**

(a) Raman G' spectra for Bernal-stacked 3-6 layer graphene whose number of layers was calibrated by using AFM. (b) Map of Raman G band signal for sample with a single gate electrode. (c) Raman spectra of the same sample.

Figure 8

**Comparison of Landau fan diagrams.**

(a) Map of magnetoresistance for the sample with only a back gate electrode. (b) Comparison of the positions of some of the energy gaps. The solid line indicates the positions of the energy gaps. Numbers are filling factors. The left panel is for the sample with a single gate electrode. Right panel is for the sample with the dual gate electrode described in the main text.

Figure 9
**Definition of Slonczewski-Weis-McClure (SWMcC) parameters.**
Open circles indicate carbon atoms at A-site in the honeycomb lattice. Filled circles indicate carbon atoms at B-site.

Figure 10
**Screening of gate-induced carriers.**
Schematic diagram of induced charges in each graphene layer due to gate voltage from top or back gate electrode. $d$ is the interlayer distance of multilayer graphene. $\lambda$ is the screening length. $\vec{E_1}, \vec{E_2}, \ldots, \vec{E_5}$ are the electric fields between the adjacent layers.

Fig1

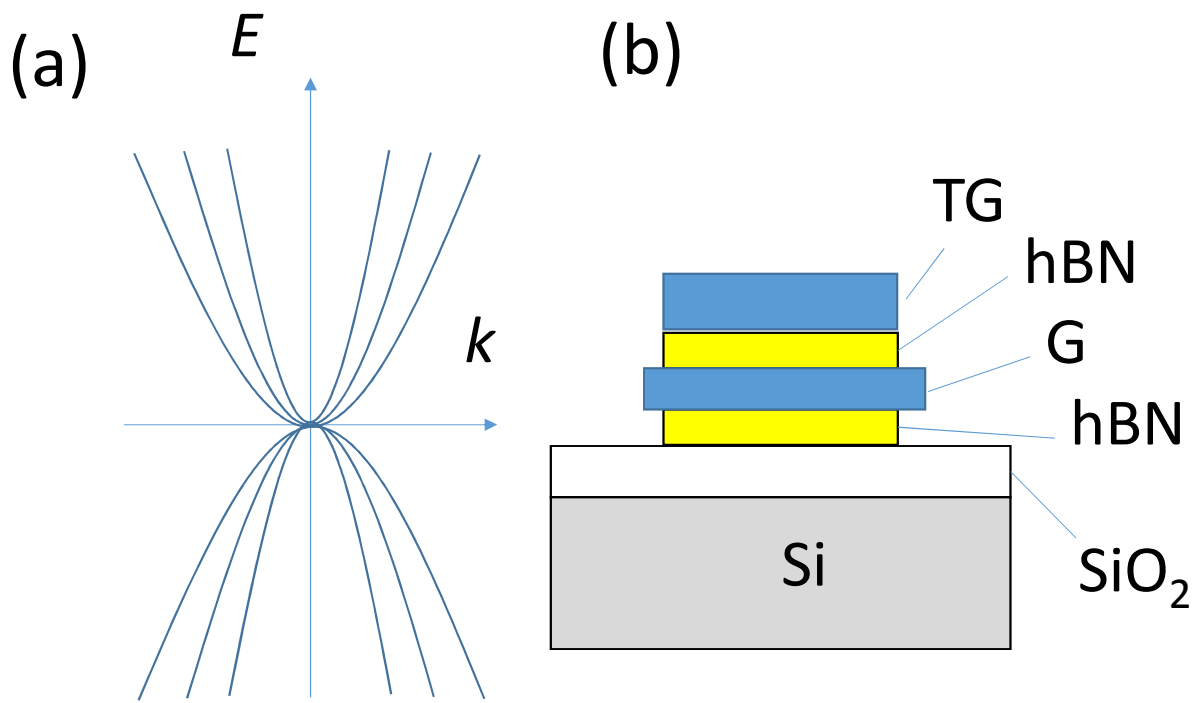

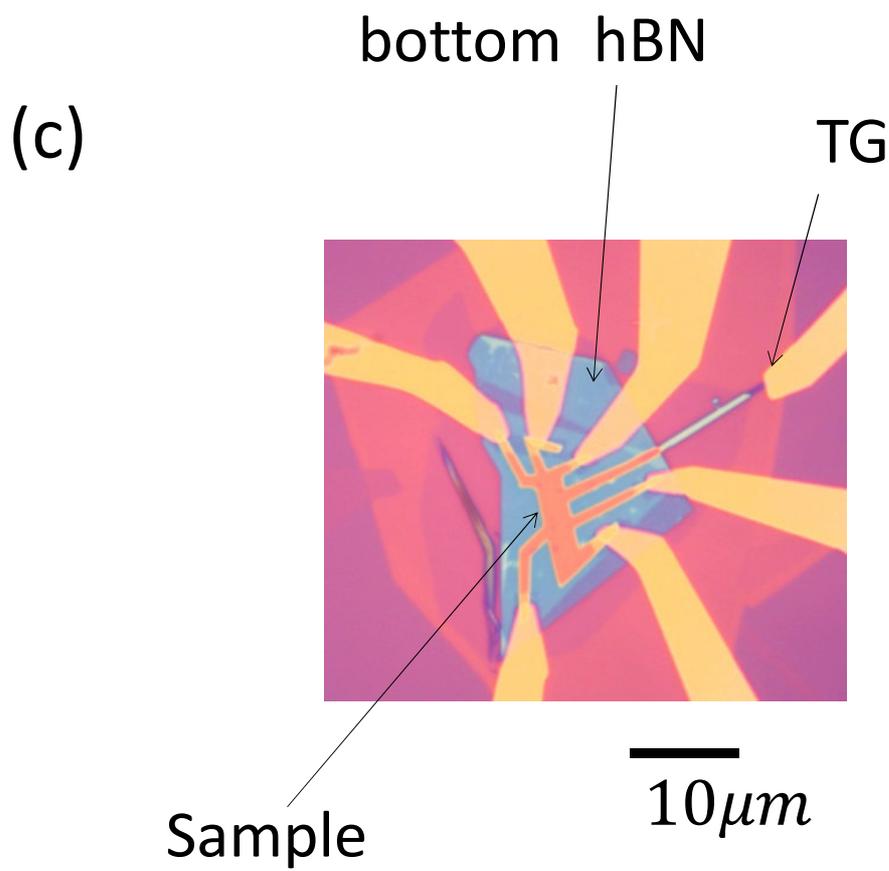

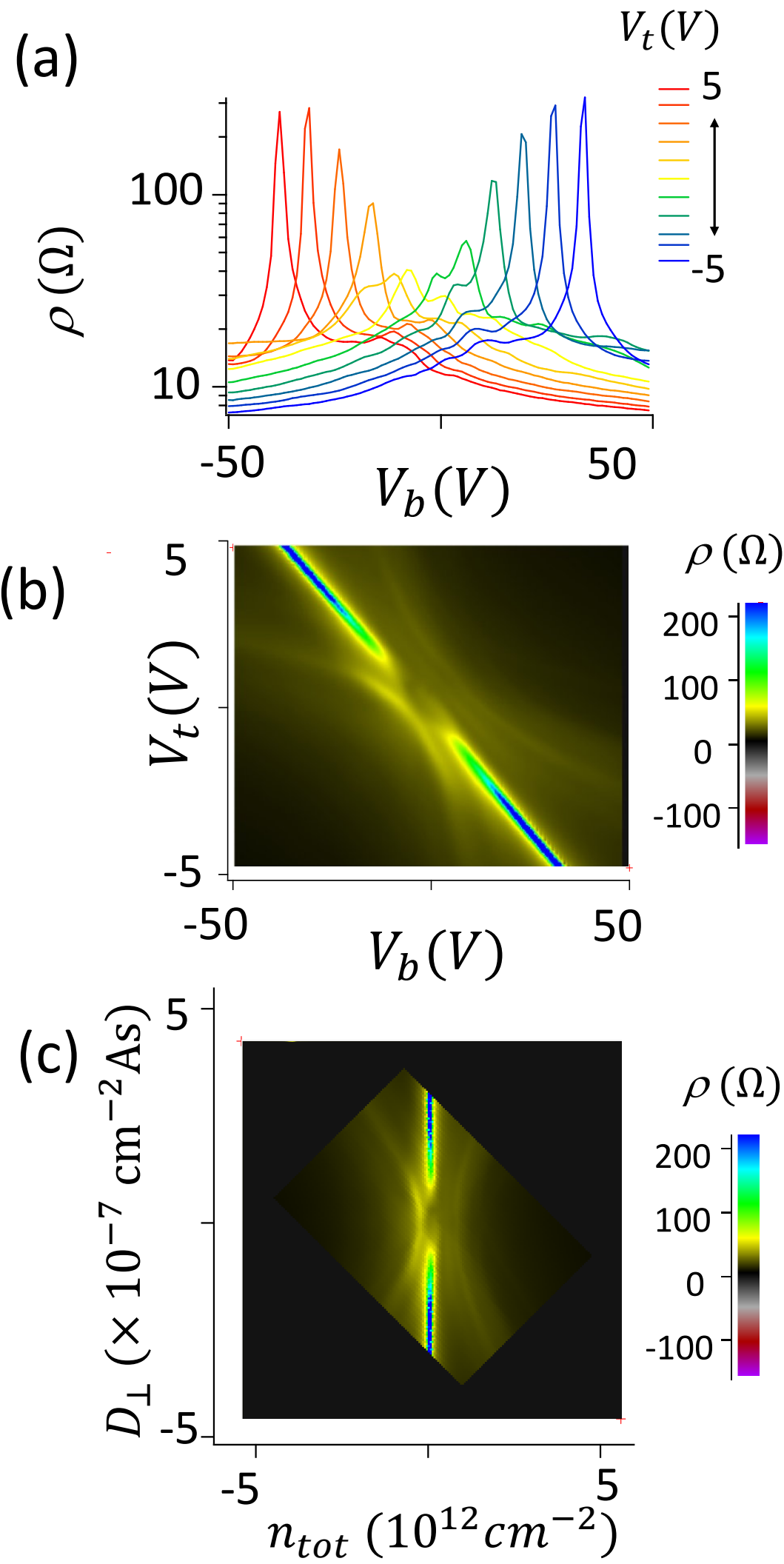

Fig 2

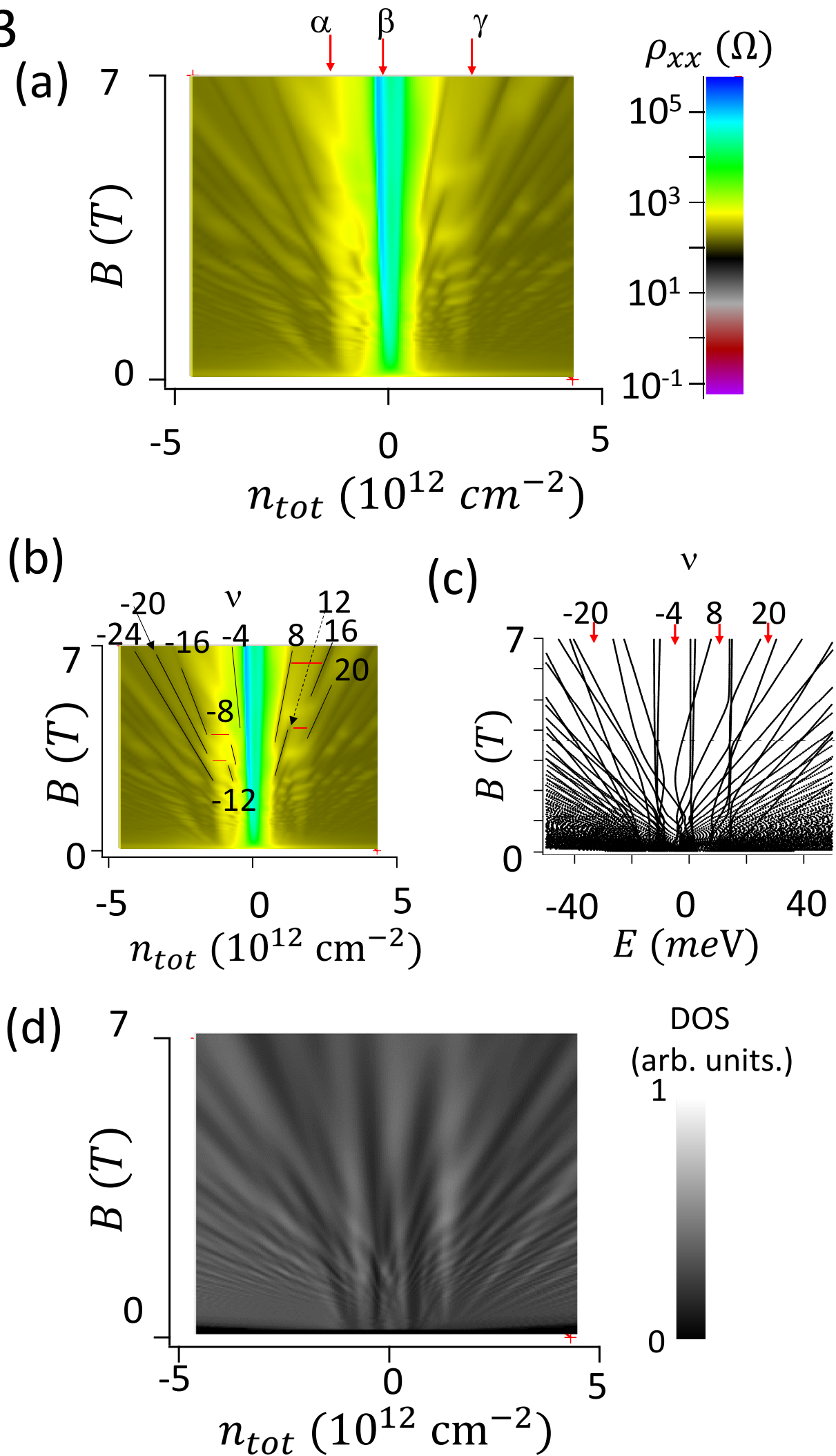

Fig. 4

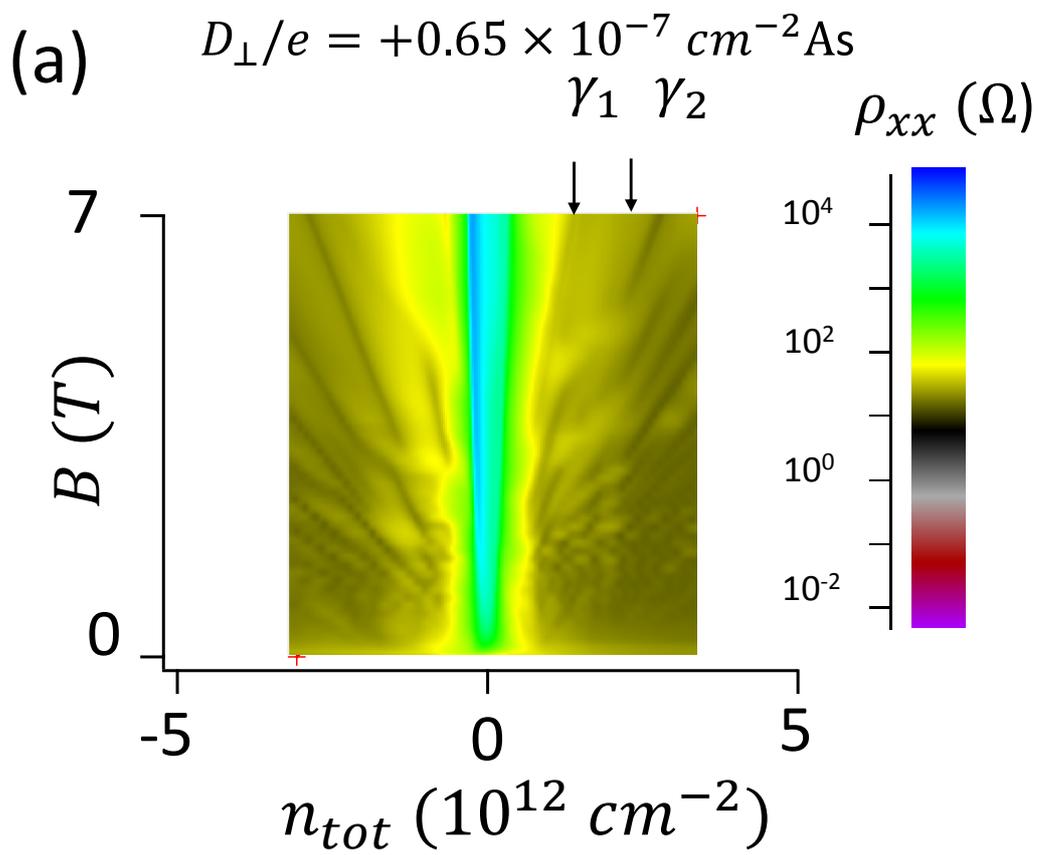

(a) $D_\perp/e = +0.65 \times 10^{-7}\, cm^{-2}\text{As}$

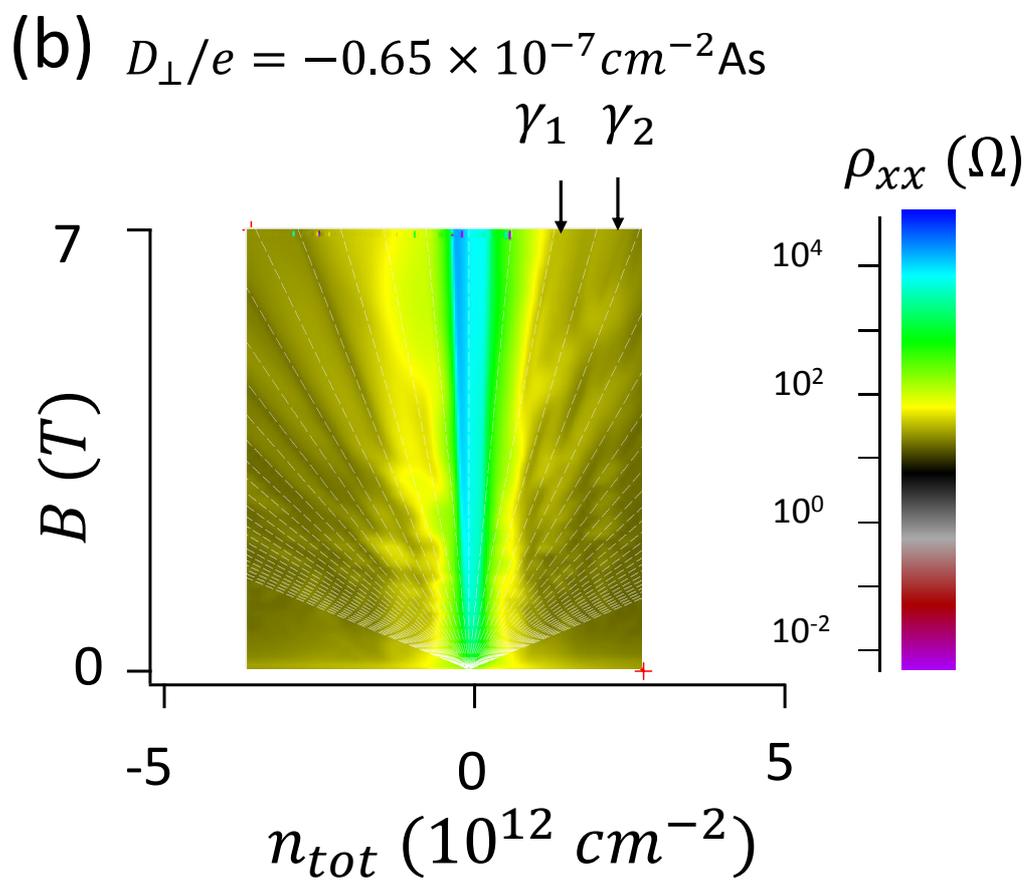

(b) $D_\perp/e = -0.65 \times 10^{-7}\, cm^{-2}\text{As}$

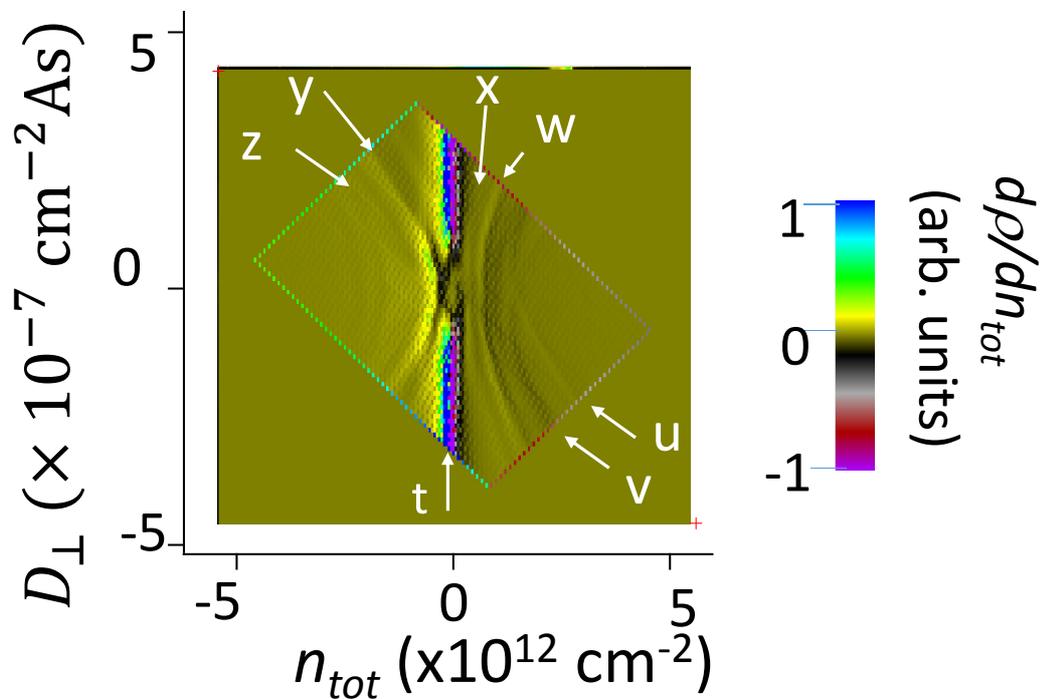
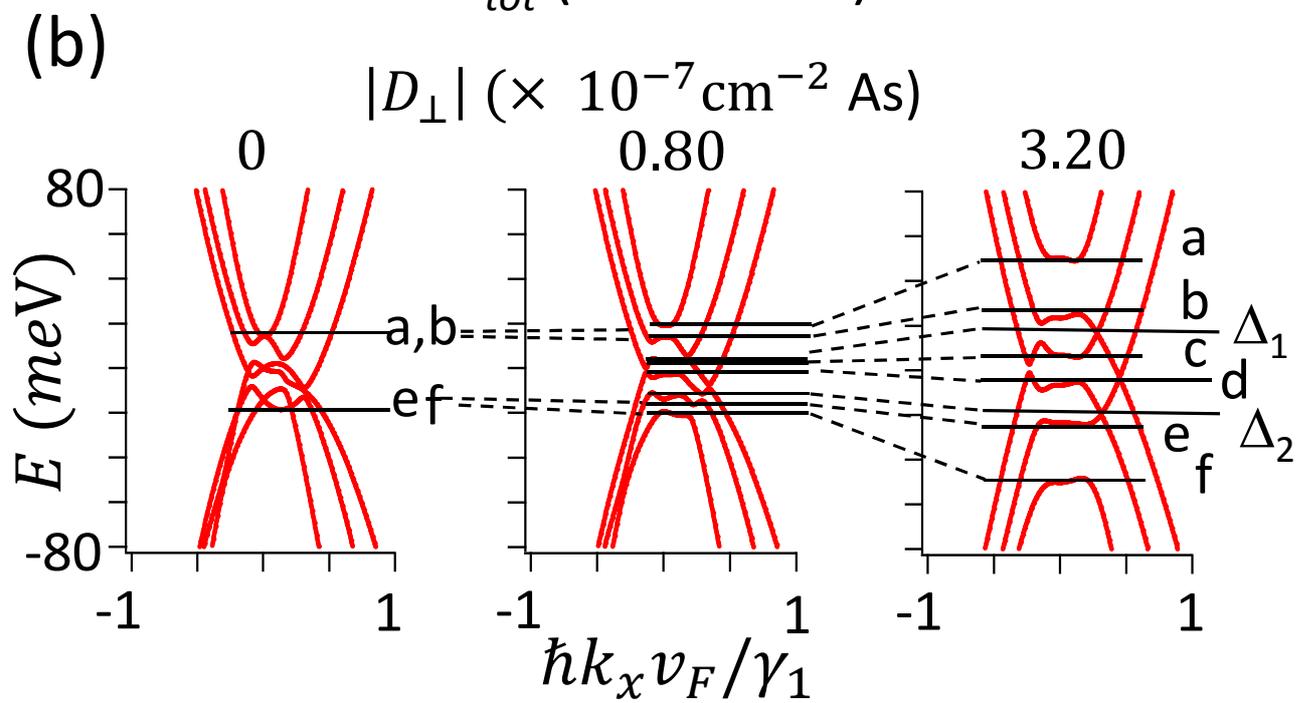
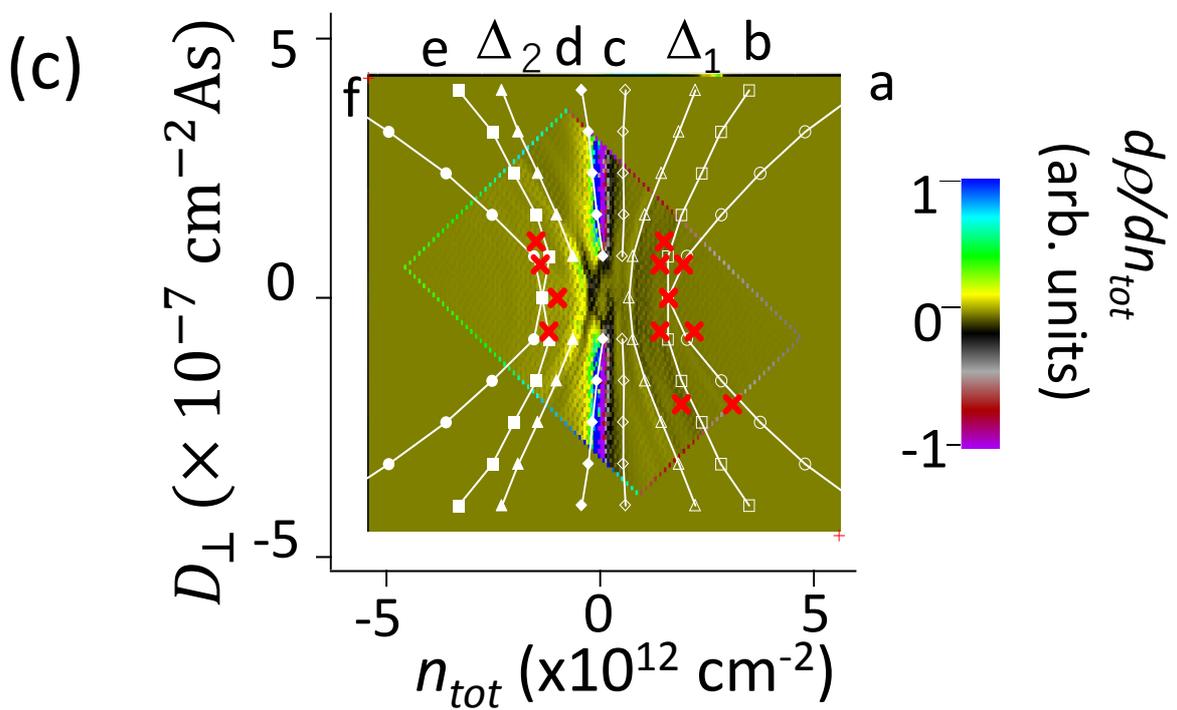

Fig. 5 (a), (b), (c)

Fig. 6

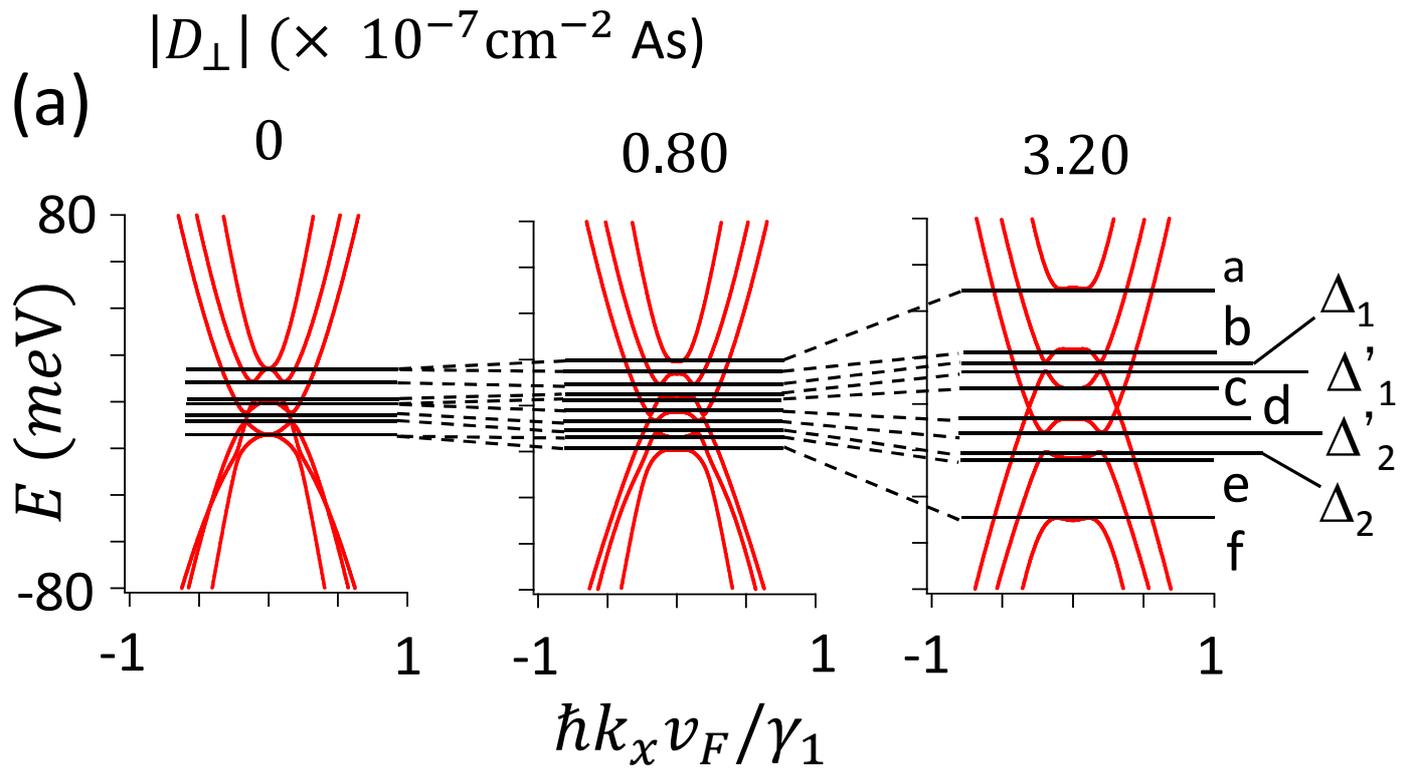

(a) $|D_\perp|$ (× $10^{-7}$ cm$^{-2}$ As)

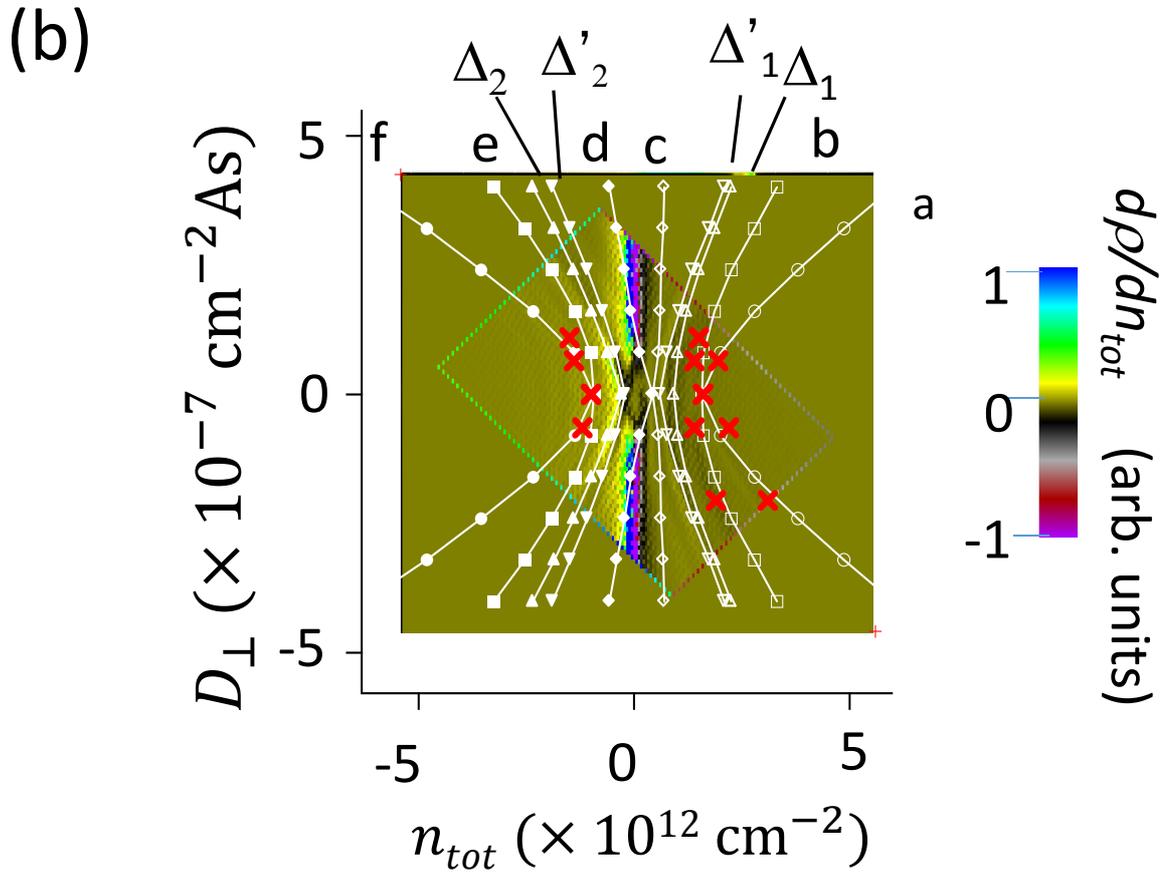

(b)

Fig. 7

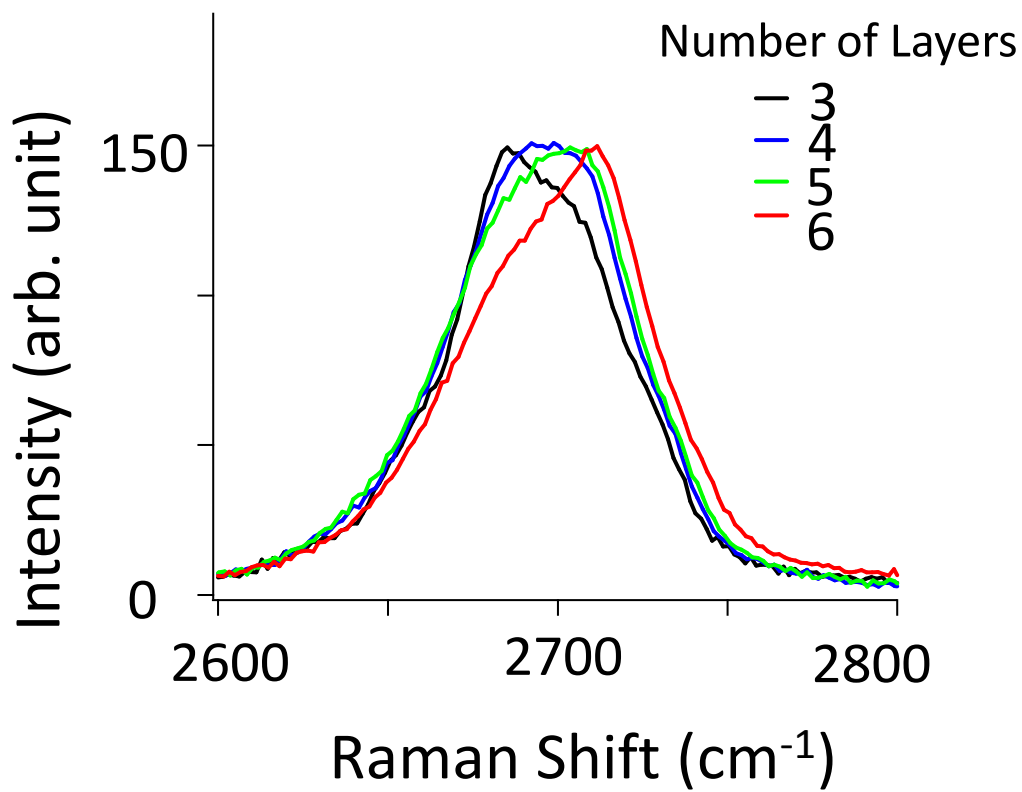

(a)

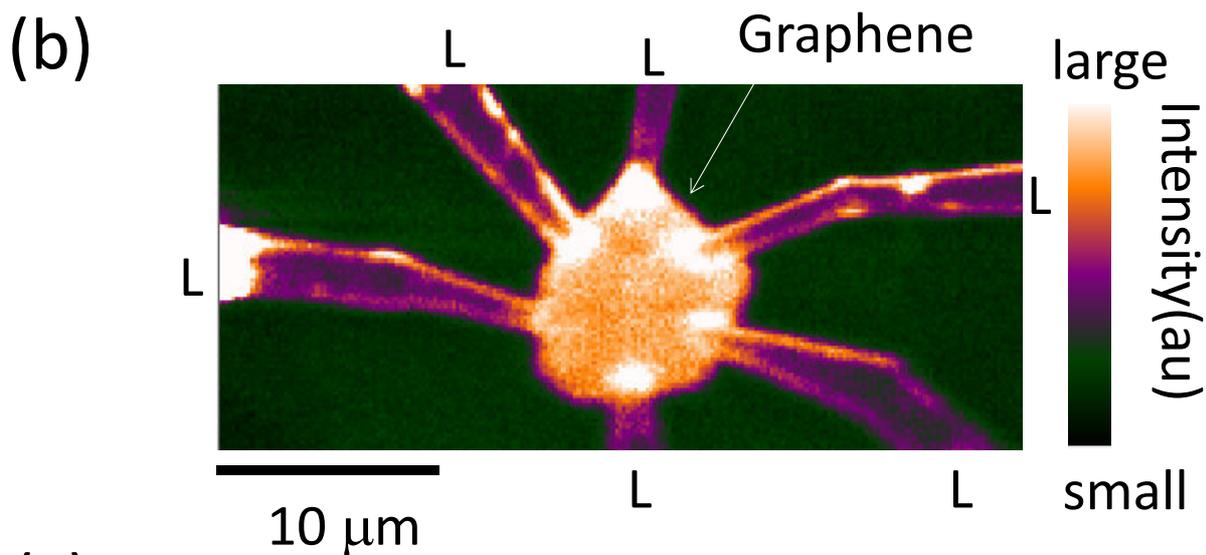

(b)

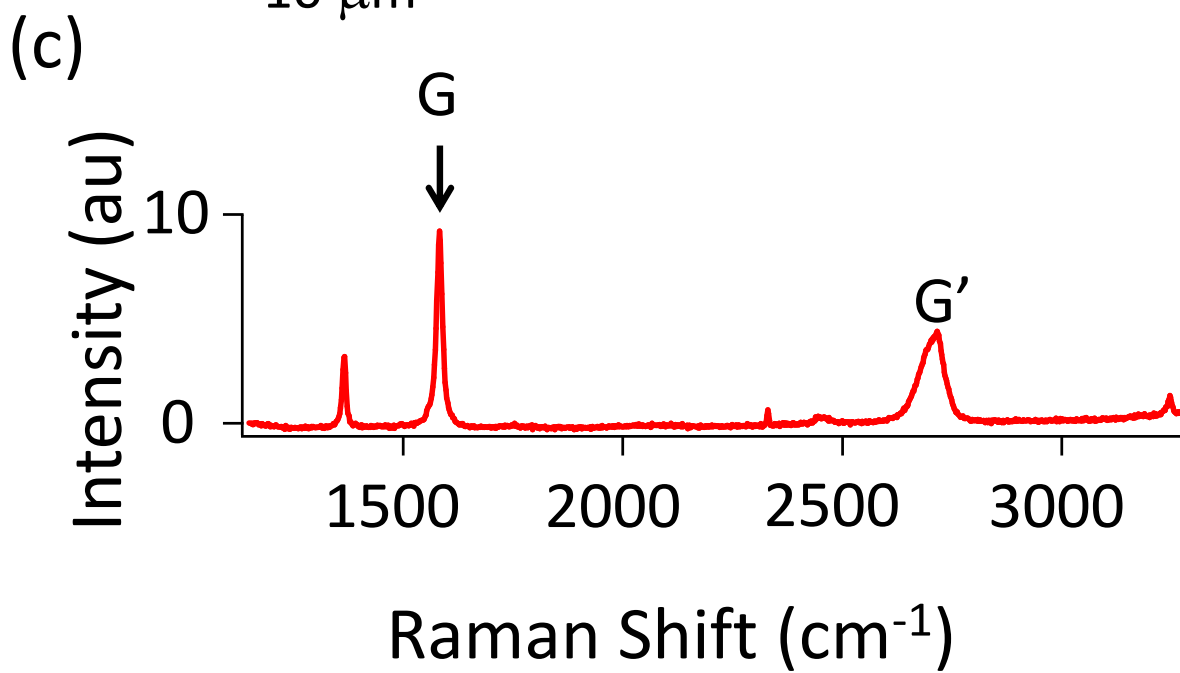

(c)

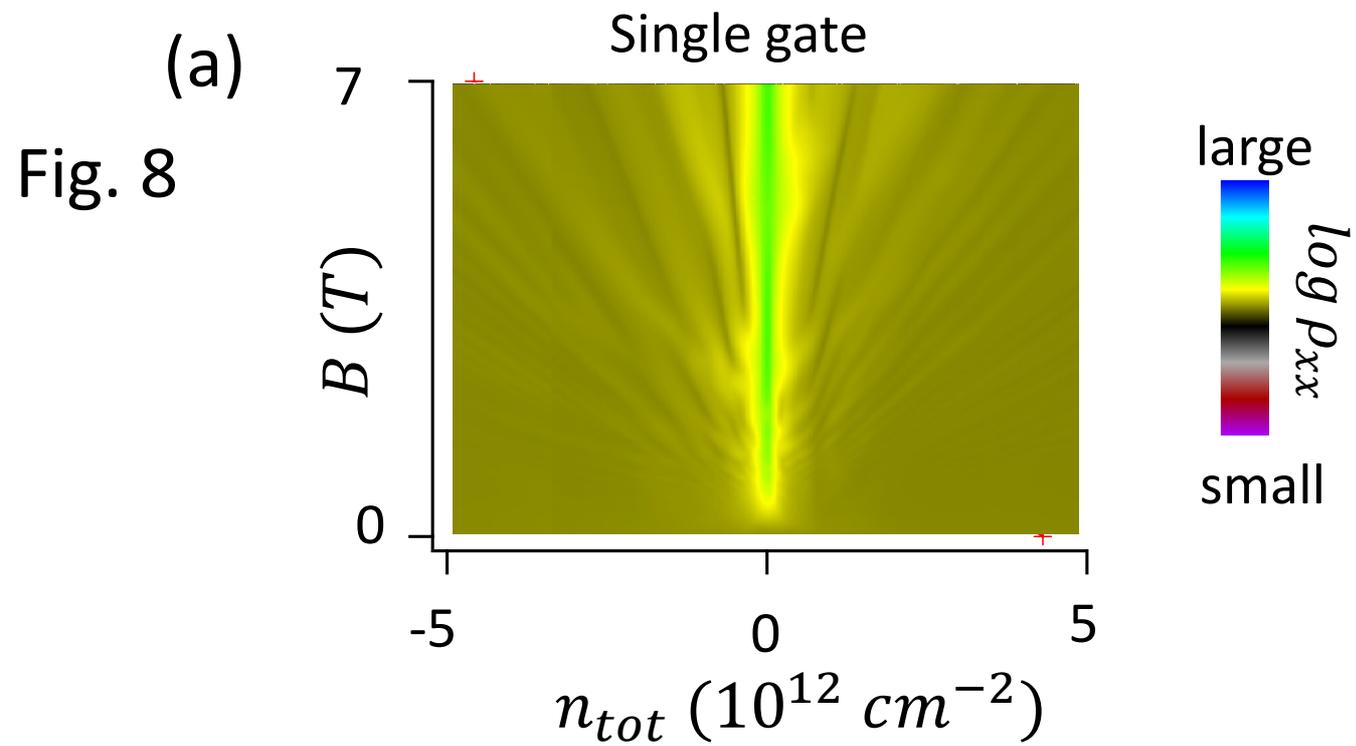

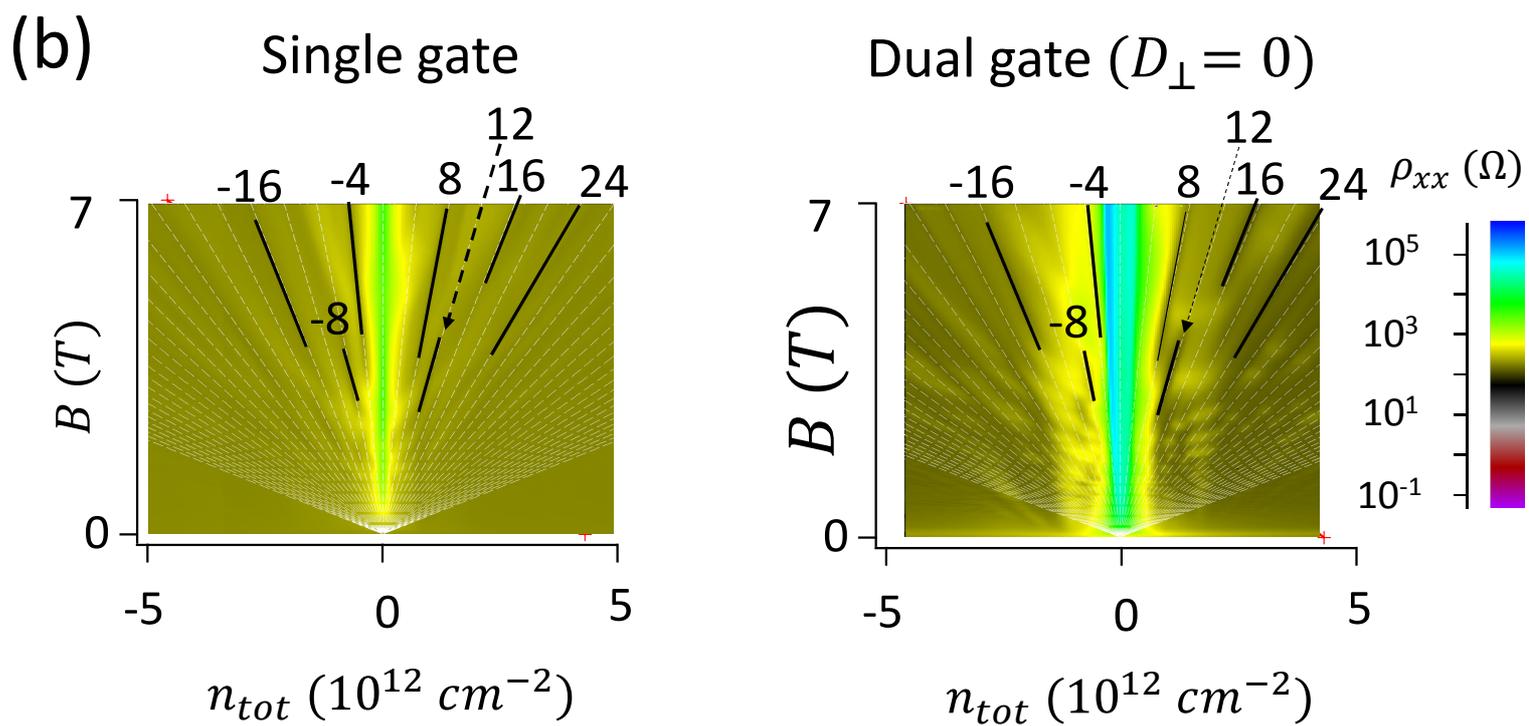

Fig. 8

Fig. 9

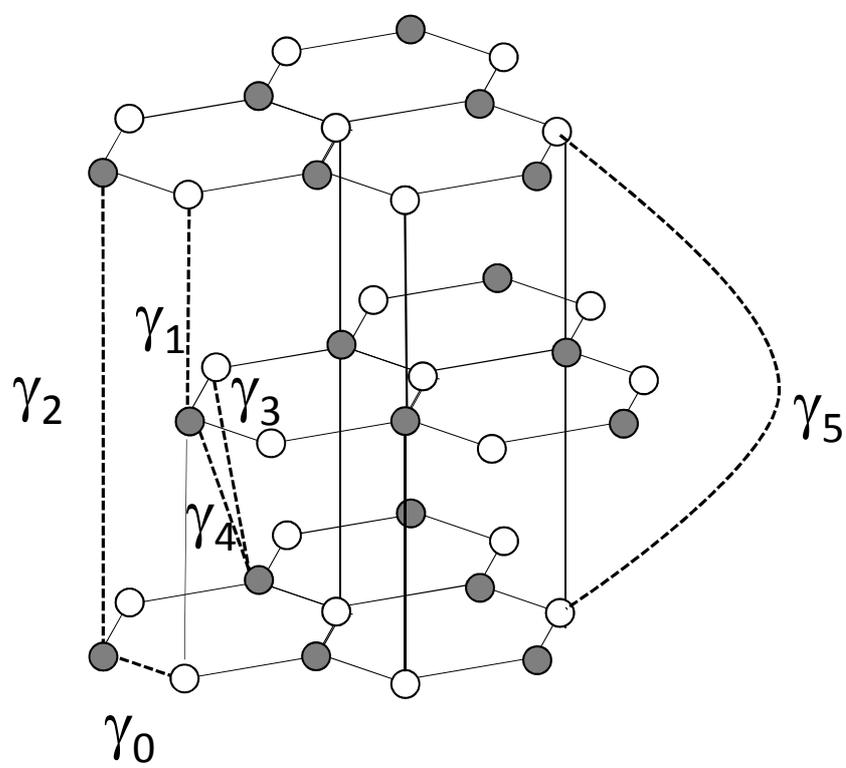

Fig. 10

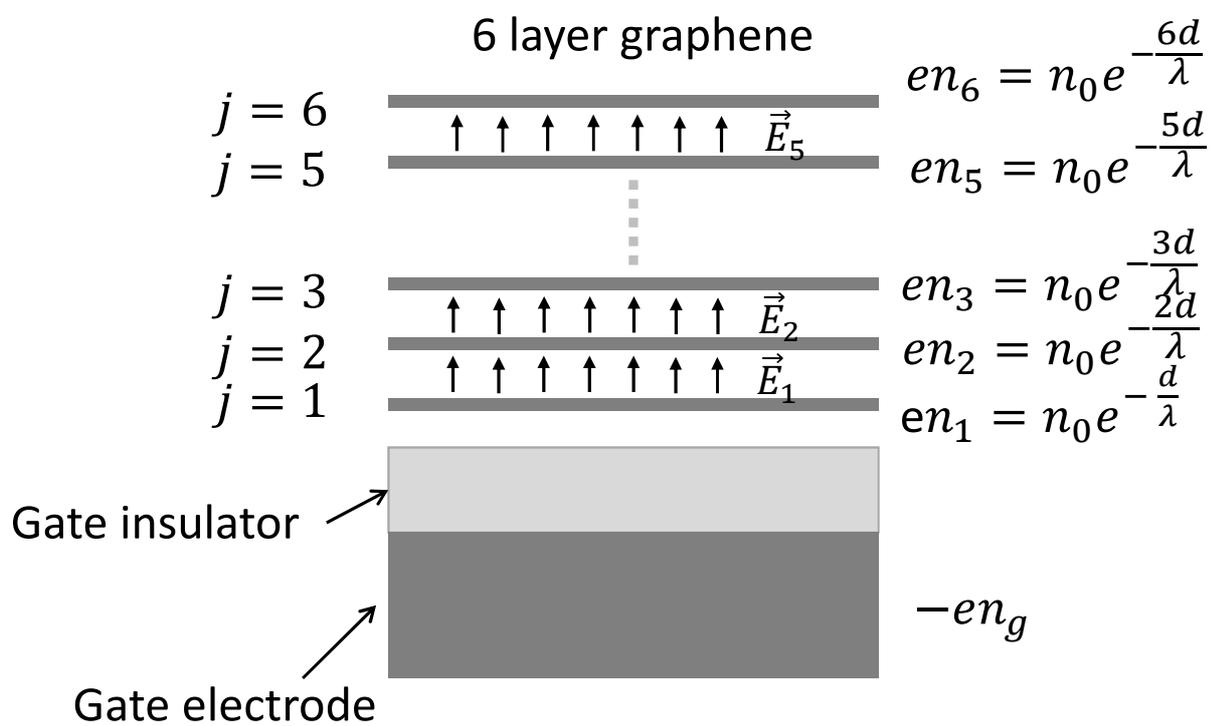